\documentclass[letterpaper, 10 pt, conference]{ieeeconf}  

\IEEEoverridecommandlockouts                              %
\usepackage[dvipsnames]{xcolor}
\usepackage{mathtools}

\title{\LARGE \bf
Prosumers Participation in Markets:\\
A Scalar-Parameterized Function Bidding Approach}

\author{Abdullah Alawad$^{1}$, Muhammad Aneeq uz Zaman$^{1}$, Khaled Alshehri$^{2}$ and Tamer Başar$^{1}$
\thanks{$^{1}$The authors are affiliated with the Coordinated Science Laboratory, University of Illinois Urbana–Champaign, Urbana, IL 61801.
        {\tt\small aalawad2@Illinois.edu, mazaman2@illinois.edu, basar1@illinois.edu}}%
\vspace{-0.4cm}
\thanks{$^{2}$ Khaled Alshehri is with the Control and Instrumentation Engineering Department and the Interdisciplinary Research Center for Smart Mobility and Logistics, King Fahd University of Petroleum and Minerals Dhahran 31261, Saudi Arabia.
        {\tt\small kalshehri@kfupm.edu.sa}}%
}
\usepackage{comment}
\usepackage{amsmath,amssymb, amsthm,bm}
\usepackage{xcolor}
\usepackage{breqn}

\usepackage{titlesec}
\titlespacing\section{0pt}{6pt plus 0pt minus 0pt}{4pt plus 0pt minus 0pt}
\titlespacing\subsection{0pt}{2pt plus 0pt minus 0pt}{6pt plus 0pt minus 0pt}
\titlespacing\subsubsection{0pt}{12pt plus 4pt minus 2pt}{0pt plus 2pt minus 2pt}

\newcommand\scalemath[2]{\scalebox{#1}{\mbox{\ensuremath{\displaystyle #2}}}}
\usepackage{scalerel}

\newcommand{\beq}{\begin{equation}}
\newcommand{\eeq}{\end{equation}}
\newcommand{\beqa}{\begin{eqnarray}}
\newcommand{\eeqa}{\end{eqnarray}}
\newcommand{\beqan}{\begin{eqnarray*}}
\newcommand{\eeqan}{\end{eqnarray*}}

\renewcommand{\[}{\left[}

\newcounter{l1}
\newcounter{l2}
\newcounter{l3}
\newcommand{\bdotlist}{\begin{list}{$\bullet$}{}}
\newcommand{\bboxlist}{\begin{list}{$\Box$}{}}
\newcommand{\bbboxlist}{\begin{list}{\raisebox{.005in}{{\tiny
$\blacksquare$ \ \ }}}{}}
\newcommand{\bdashlist}{\begin{list}{$-$}{} }
\newcommand{\blist}{\begin{list}{}{} }
\newcommand{\barablist}{\begin{list}{\arabic{l1}}{\usecounter{l1}}}
\newcommand{\balphlist}{\begin{list}{(\alph{l2})}{\usecounter{l2}}}
\newcommand{\bAlphlist}{\begin{list}{\Alph{l2}.}{\usecounter{l2}}}
\newcommand{\bdiamlist}{\begin{list}{$\diamond$}{}}
\newcommand{\bromalist}{\begin{list}{(\roman{l3})}{\usecounter{l3}}}

\newtheorem{theorem}{Theorem}
\newtheorem{lemma}{Lemma}

\newtheorem{assumption}{Assumption}

\begin{document}
\maketitle
\thispagestyle{empty}
\pagestyle{empty}

\begin{abstract}
In uniform-price markets, suppliers compete to supply a resource to consumers, resulting in a single market price determined by their competition. For sufficient flexibility, producers and consumers prefer to commit to a function as their strategies, indicating their preferred quantity at any given market price. Producers and consumers may wish to act as both, i.e., prosumers. In this paper, we examine the behavior of profit-maximizing prosumers in a uniform-price market for resource allocation with the objective of maximizing the social welfare. We propose a scalar-parameterized function bidding mechanism for the prosumers, in which we establish the existence and uniqueness of Nash equilibrium. Furthermore, we provide an efficient way to compute the Nash equilibrium through the computation of the market allocation at the Nash equilibrium. Finally, we present a case study to illustrate the welfare loss under different variations of market parameters, such as the market's supply capacity and inelastic demand.
\end{abstract}

\section{Introduction} \label{section 1}
The competition for a divisible resource between selfish agents has made game-theoretic methods useful tools for the design of resource allocation mechanisms. For such mechanisms design, several metrics have been investigated in the literature, such as fairness and social welfare efficiency of the allocation as well as the computational cost for finding the allocation where strategy space plays a significant role. One measure of fairness was discussed in \cite{kelly1998rate} for a proportionally fair (PF) pricing mechanism where the resulting allocation makes it impossible to increase the sum of weighted proportional gains. Another design metric is the efficiency of allocation with respect to social welfare maximization, i.e., to what extent the sum of agents’ utilities is close to the maximum possible value. The efficiency of the aforementioned PF pricing mechanism, however, is undermined when agents behave strategically, i.e., when their strategies incorporate the relationship of the price to the bids---this turns the mechanism into an auction. In a competitive formulation, \cite{johari2004efficiency} studied the efficiency loss of this PF auction and showed that it is 25\% in the worst case. 

Vickrey-Clarke-Groves (VCG) is a well-known class of mechanisms \cite{vickrey1961counterspeculation,clarke1971multipart,groves1973incentives} for resource allocation which ensures that truthful reporting of each agent is a dominant strategy. However, it may not be practical for some domains because of shortcomings such as providing a different price to different agents for the same resource. Other mechanisms similar to VCG were studied for pricing divisible resources with scalar strategy spaces such as \cite{maheswaran2003nash,maheswaran2006efficient}; see section 2 of \cite{johari2011parameterized} for an extended list. \cite{maheswaran2003nash} investigated a PF divisible auction in which the notions of price and demand functions were introduced to characterize optimal response functions of the agents. A unique Nash equilibrium was proven to exist for agents with heterogeneous quasilinear utilities and a decentralized iterative algorithm was described to converge to the equilibrium. A class of mechanisms with single-dimensional signaling (bidding strategy) was studied in \cite{maheswaran2006efficient} such that the PF auction was shown to be inefficient in general. An infinite subclass of the PF auctions called efficient signal proportional allocation (ESPA) mechanisms was shown to maximize the social welfare for agents with quasi-linear utilities. Besides efficiency and due to the one-dimensional signaling space, computational cost is another design metric in which ESPA is an optimal allocation mechanism.
\vspace{-0.05cm}
\subsection{Bidding strategy: towards scalar-parameterized functions} \label{section 1A}
\vspace{-0.17cm}
The analysis of supply function equilibria (SFE) for uniform-price markets is closely related to this growing literature on efficiency guarantees in market design. In such bidding mechanisms and for sufficient flexibility, competing suppliers prefer to commit to (offer) supply functions as their strategies, indicating their preferred supply quantity at any given market price. This is in contrast to committing to a scalar strategy, such as a fixed price (Bertrand model) or a fixed quantity (Cournot model). \cite{klemperer1989supply} investigated the existence of Nash equilibria resulting from supply function offers and demonstrated that they can be highly inefficient. 

In a centralized uniform-price market-clearing mechanism for supply-quantity allocation of an infinitely divisible resource, \cite{johari2011parameterized} proposed a restriction on the class of supply functions, limiting the supplier's strategy to scalar-parameterized functions. Under a fixed, inflexible (inelastic) demand and suppliers with maximum production capacity, the paper studied the existence of Nash equilibrium and the efficiency of its associated market allocation. This formulation was extended in \cite{lin2019structural} to study several inelastic demands, besides the strategic suppliers, that are spread throughout a transmission-constrained power network. By studying the efficiency of Nash equilibrium's market allocation, the paper explained how market share and residual supply index can predict the extent to which suppliers can exert market power to influence the market outcome to their advantage. \cite{ndrio2020scalar} extended the formulation to study two-sided markets (strategic consumers and producers) in which the participants' strategies are scalar-parameterized functions where the consumers have both elastic and inelastic (minimum) demands. In some markets, the participants may wish to act as both, producers and consumers, in which they are often called prosumers. The formulation in \cite{ndrio2020scalar} can be used to serve this purpose where each prosumer acts simultaneously as a producer and a consumer, maximizing two separate payoff functions and submitting two scalar-parameterized functions. However, the two payoff functions might be conflicting for the prosumer. This paper proposes a bidding mechanism for the prosumers in which they maximize a single payoff function.
\vspace{-0.05cm}
\subsection{Prosumer markets: peer-to-peer and other typologies}\label{section 1B}
\vspace{-0.17cm}
A prime example of an infinitely divisible resource with ample research on prosumer markets is electricity. Different typologies for prosumer electricity markets were discussed in \cite{parag2016electricity}. From a game-theoretic point of view, the discussed typologies can be classified into three main prosumer market models: peer-to-peer model in which competitive prosumers exchange resources with each other (e.g. within an interconnected neighborhood), prosumer-to-aggregators model in which competitive prosumers exchange resources with a central aggregator (e.g. microgrid-operator) which may also exchange with other aggregators and/or an upstream market (e.g. wholesale market), and organized-prosumer-group model in which a group of prosumers cooperate to form a bulk supplier/prosumer (e.g. virtual power plant) which exchanges resources with an upstream market.

In competitive settings, whether the typology of prosumers is peer-to-peer or prosumer-to-aggregators, several solution concepts in game theory are useful for analyzing such typologies \cite{saad2012game}. For example, Nash equilibrium is suitable for non-cooperative games where no player dominates the decision process whereas Stackelberg equilibrium is useful when hierarchy is allowed in the decision process \cite{bacsar1998dynamic}. 

Compared to bulk producers in the wholesale electricity markets, the small size of prosumers often prevents them from participating in such markets. Hence, retail aggregators offer a reasonable solution to enable the participation of prosumers in wholesale markets. In this trading setting where a profit-maximizing retail agreggator sets a uniform price for its competitive prosumers, \cite{alshehri2020quantifying} formulated a Stackelberg game between the aggregator and the prosumers and characterized the Stackelberg equilibrium. The goal was to quantify the loss in efficiency (e.g. welfare loss) that may result from the strategic incentive of the aggregator, when compared to the benchmark efficiency in which the prosumers directly participate in the wholesale market. Another aggregator framework was introduced in \cite{gharesifard2015price} where the aggregator incentivizes its prosumers to produce or consume energy over a period of time by setting two prices: one for production and another for consumption. Focusing on the prosumer's strategic decision-making process, the paper established sufficient conditions on the aggregator's pricing strategy for the existence of a unique Nash equilibrium for a game formulated for the prosumers with the aggregator.

Controllable-loads and interconnected microgrids are other agents similar to prosumers in their involvement in distributed decision-making in electricity markets \cite{saad2012game}. The aforementioned scalar-parameterized supply functions were reformulated in \cite{xu2015demand} to serve as a bidding mechanism for consumers equipped with controllable-loads in a peer-to-peer setting. The paper studied the efficiency of the allocation of consumers' load adjustment capacities at the Nash equilibrium. However, since the objective function of a controllable-demand agent does not include production cost, this agent is more similar to pure-consumer agents than prosumers and interconnected microgrids who are equipped with production capacity (directly or by proxy).
\vspace{-0.05cm}
\subsection{Mechanism design objectives and prosumer nature}\label{section 1C}
\vspace{-0.17cm}
Among the several typologies above for the prosumer markets, this paper restricts the analysis to a peer-to-peer model. In the design of a uniform-price market where strategic (price-anticipating) participants compete for a divisible resource, we generally seek to have an efficient Nash equilibrium. In other words, we want the market's aggregate cost at the Nash equilibrium to be close to the minimum possible cost (i.e. cost at the price-taking competitive equilibrium or the socially-optimal cost). Equivalently, we seek to have a social welfare at the Nash equilibrium that is close to the socially-optimal welfare. Following the line of research on scalar-parameterized function bidding, our goal in this paper is to study the market design question of how to formulate a scalar-parameterized function bidding that  provides sufficient flexibility for the prosumers to declare their bidding preferences in a way that yields an efficient allocation of productions and consumptions, minimizing the ``welfare loss'' that occurs due to their strategic behavior. 

Due to its ability to simultaneously produce and consume (i.e. prosumer duality), the prosumer may choose to meet (part of) its demand by importing rather than utilizing its production capacity first and/or to utilize (part of) its production capacity for exporting rather than consuming it locally. Therefore, the prosumer may always have consumption cost and production revenue. By expanding the traditional Cournot model, \cite{tsybina2022effect} demonstrated via simulations that prosumer duality can lead to more competitive behavior than pure producers in a traditional producer/consumer system. In other words, the prosumers' best-response supply quantities are closer to the competitive levels than those of the traditional producers, under the same game-theoretic scenario. The design that we propose in this paper, however, does not account for prosumer's duality and assumes that positive supply can only occur after meeting the local demand.

The rest of the paper is organized as follows: Section \ref{section 2} introduces the bidding mechanism for the prosumers and the market model (optimization of resource allocation). Section \ref{section 3} discusses the competitive equilibrium where the prosumers are price-takers. Section \ref{section 4} investigates the Nash equilibrium where the prosumers are price-makers (strategic). Section \ref{section 5} outlines a case study where the market outcome (allocation) is examined with respect to the welfare loss when market's supply capacity or inelastic demand are varied. Section \ref{section 6} concludes and provides future directions. 

\section{Bidding Mechanism and Market Model} \label{section 2}
To investigate the posed market design question, we consider profit maximizing prosumers having production costs and utilities characterized, respectively, by convex cost functions and concave utility functions in the output quantity. Each prosumer has a maximum production capacity and a minimum inelastic demand. Analogous to the scalar-parameterized function proposed as a bidding mechanism for the producer in \cite{johari2011parameterized} and for the producer and the consumer in \cite{ndrio2020scalar}, our goal is to propose a scalar-parameterized function for the prosumer. As mentioned in Section \ref{section 1A}, it is straightforward to see that the two scalar-parameterized functions proposed in \cite{ndrio2020scalar} can be utilized for the prosumer case. In this setting, the prosumer effectively acts as two agents, maximizing two (possibly conflicting) payoff functions separately and submitting both scalar-parameterized functions. We can eliminate this possible conflict by having the prosumer optimize a single objective function. Using the formulation in \cite{ndrio2020scalar}, one might conjecture that having the prosumer maximize a single payoff function defined by the summation of the two payoff functions for the producer and the consumer, using the same two decision variables, results in a game with a Nash equilibrium. However, we briefly highlight that this is not true. While a competitive equilibrium exists when the prosumers are price-takers (i.e. perfect competition), a Nash equilibrium cannot exist when the prosumers are price-anticipating (strategic). This is because the resulting payoff function of the strategic prosumer is not concave in its two decision variables, hence it may increase without a bound. In contrast, the formulation we propose involves a single decision variable for the prosumer's payoff function. Besides eliminating the possible conflict of maximizing two separate payoff functions, utilizing a single scalar-parameterized function, in contrast to utilizing two, would reduce the computation cost for the market operator (central clearinghouse) when solving the optimization of resource allocation. 

Denote the set of prosumers by $\mathcal{N}$=$\{1,2,\ldots,N\}$. Let $q_i \in \mathbb R$ denote the desired quantity of demand (positive) or supply (negative) for each prosumer $i$, $d_{min}\in \mathbb{R}^+$ represent the minimum inelastic (inflexible) demand for each prosumer, and $s_{max}\in \mathbb{R}^+$ denote the maximum supply capacity. Their values are assumed to be identical for all prosumers without loss of generality. Throughout this paper, we emphasize the distinction between the supply capacity $s_{max}$ and the production capacity. The production capacity includes not only the supply capacity but also the capacity used by each prosumer to meet its own demand, i.e., the production capacity is $s_{max}+d_{min}$. In addition, $p\in \mathbb{R}^+$ is the uniform-price to be determined by the market operator to clear the market. We propose the following bidding mechanism which consists of two parts: a scalar-parameterized function which gives the quantity $q_i$, and the scalar $s_{max}$ introduced earlier:
\vspace{-2pt}
\begin{equation}\label{eq6}
    q_i = Q(\theta_i,p) \coloneqq d_{min} + \frac{\theta_i}{p}, \ \ \text{and} \ s_{max}, \ \ i \in \mathcal{N}
    \vspace{-2pt}
\end{equation}
Note that $d_{min}$ is included in the scalar-parameterized function. Also, $q_i$ and $s_{max}$ in the bidding mechanism \eqref{eq6} must satisfy $s_{max} \ge -q_i$. Prosumer $i$ chooses the parameter $\theta_i{\in}\mathbb R$ such that $\theta_i{>}0$, $-pd_{min}{\le}\theta_i{\leq}0$, and $\theta_i{<}-pd_{min}$ represent the pure-consumption mode (i.e. imported consumption with no production), the prosumption mode (i.e. imported consumption with production only for local demand), and the pure-supply mode (i.e. production with no imported consumption). Thus for each prosumer $i$, $q_i{>}d_{min}$ is the pure-consumption mode such that $q_i$ is the “total consumption” quantity consumed from the market, $0\leq q_i \leq d_{min}$ is the prosumption mode such that $q_i$ is the “imported consumption” quantity consumed from the market (i.e. the prosumer imports $q_i$ and produces $d_{min} - q_i$), and $q_i{<0}$ is the pure-supply mode such that $|q_i|$ is the “supply” quantity supplied to the market (i.e. the prosumer produces $|q_i|+d_{min}$, of which $|q_i|$ is supplied to the market). Note that the inelastic demand $d_{min}$ is satisfied in all modes, however, from different sources: entirely from the market in the pure-consumption mode, partially from the local production $d_{min}-q_i$ in the prosumer mode, and entirely from the local production $d_{min}$ in the supply mode. Note also that using this bidding mechanism, each prosumer cannot supply to other prosumers until it produces its entire inelastic demand, i.e., the mechanism does not enjoy the duality of prosumers. 

From \eqref{eq6}, we note that in the prosumption mode, $q_i{>}0$ (imported consumption quantity) increases as $p$ (the price) increases, and in the supply mode, $|q_i{<}0|$ (supply quantity) decreases as $p$ increases. The function in both regions does not follow the typical demand (supply) curve for the consumer (producer), i.e., the preference curves reflecting that with higher prices consumers want to consume less and producers want to produce more. However, we will see that given the payoff function \eqref{eq11} that the prosumer aims to maximize, the prosumer may follow the aforementioned typical preferences because the optimal quantity of $q_i$ depends not only on the value of $p q_i$ but also on the cost/utility function $S_i(q_i)$. This means that the prosumer should define $S_i(q_i)$ taking into account the atypical demand/supply curves so that the payoff function \eqref{eq11} results in optimal imported consumption quantity (supply quantity) that is smaller (larger) with higher prices.

The market operator solves the following convex optimization problem to maximize the aggregate social welfare defined in \eqref{eq7a}:
\vspace{-15pt}
\begin{subequations}\label{eq7}
\begin{alignat}{2}
    & \underset{\boldsymbol{q}} {\text{maximize}} \ \ \mathcal{W}(\boldsymbol{q})\coloneqq \ \ &&  \sum_{i=1}^N  S_i(q_i), \label{eq7a} \\
    & \text{subject to} &&  \sum_{i=1}^N q_i = 0, \label{eq7b} \\
    &&&  s_{max} \ge -q_i, \ \forall i \in \mathcal{N} \label{eq7c} 
\end{alignat}
\end{subequations}
where $S_i(q_i)$ is the utility/cost function for prosumer $i$, i.e., utility function when it is positive or cost function when it is negative. Any solution $\boldsymbol{q}$ (i.e. allocation profile) to \eqref{eq7} is referred to as an efficient allocation. We impose the following assumption on $S_i(q_i)$:
\vspace{-2pt}
\begin{assumption}\label{assumption1} 
For $\forall i \in \mathcal{N}$, $S_i(q_i)$ is twice continuously differentiable, strictly increasing, strictly concave, and $S_i(d_{min})=0$ where $q_i \in \mathbb{R}$ and $d_{min} \in \mathbb{R}^+$.
\end{assumption}
The market operator chooses the price $p(\boldsymbol{\theta})>0$ to clear the market, i.e., so that the supply/demand balance constraint \eqref{eq7b} $\sum_{i=1}^N Q(\theta_i,p(\boldsymbol{\theta})) = 0$ is satisfied in which case:
\vspace{-3pt}
\begin{equation}\label{eq9}
    p(\boldsymbol{\theta}) = -\frac{\sum_{i=1}^N  \theta_i}{Nd_{min}}
    \vspace{-3pt}
\end{equation}
$p(\boldsymbol{\theta})\geq 0$ is only possible if $\sum_{i=1}^N  \theta_i \leq 0$ (assumed). If the latter is zero then $q_i = d_{min}$ regardless of the value of $p$.  Hence, the following conventions are adopted which make the price continuous in $\boldsymbol{\theta}$:
\vspace{-5pt}
\begin{equation}\label{eq10}
     Q(0,0)=d_{min}, \ \ \mathrm{and} \ \ p(\boldsymbol{0})=0 
     \vspace{-3pt}
\end{equation}
Due to the assumption $\sum_{j=1}^N  \theta_j \leq 0$, the action parameter $\theta_i$ for each prosumer $i \in \mathcal{N}$ must stay within $\theta_i \leq - \sum_{j\ne i}^N  \theta_j$ which is enforced by the market operator. 
\section{Perfect Competition and Competitive Equilibrium} \label{section 3}
In this section, we present the case where all prosumers are price takers and the goal is to analyze the market outcome by establishing the existence and characterization of a unique competitive market equilibrium. Therefore, we can conclude that the allocation at the competitive equilibrium is efficient, which is established by the first fundamental theorem of welfare economics. Given the market price $\mu > 0$, prosumer $i$ maximizes the following payoff function:
\vspace{-2pt}
\begin{equation}\label{eq11}
    \pi_i^p(\theta_i,\mu) = S_i(Q(\theta_i,\mu))-\mu Q(\theta_i,\mu)
    \vspace{-2pt}
\end{equation}
Based on the definition in \eqref{eq6}, let $Q(\theta_i,\mu) = q_i$ in \eqref{eq11}. When $q_i{>} d_{min}$, then $S_i(q_i){>}0$ represents the utility gained from consuming the amount $q_i$. When $q_i{<}0$, then $|S_i(q_i){<}0|$ represents the cost incurred from supplying the amount $|q_i|$ (i.e. producing the amount $|q_i|+d_{min}$). When $0{\leq} q_i{\leq} d_{min}$, then $|S_i(q_i){\leq}0|$ (assumed) represents the cost incurred from prosuming the amount $q_i$ (i.e. consuming $q_i$ from the market while producing $d_{min}-q_i$ for local consumption). Also based on whether $q_i$ is positive or negative, the second term in \eqref{eq11} represents the cost of consumption or the revenue from supply, respectively. It is worth noting that this formulation allows the payoff \eqref{eq11} to be negative, e.g., when $0{\leq} q_i{\leq} d_{min}$, both terms in \eqref{eq11} are negative. Furthermore, the optimal social welfare \eqref{eq7a} can be negative depending on the structure of the functions $S_i(q_i), i \in \mathcal{N}$; we will see in Section \ref{section 5} that the case study results in negative optimal social welfare since the values of the example functions are larger in magnitude over the negative domains than the positive counterparts. The following theorem states the result characterizing the unique competitive equilibrium, and makes a conclusion about the corresponding allocation. Appendix \ref{Pthrm1} provides the proof.
\vspace{-2pt}
\begin{theorem}\label{theorem1}
    Suppose Assumption \ref{assumption1} holds. Then, there exists a unique competitive equilibrium, i.e., a scalar $\mu$ given by \eqref{eq10} and a vector $\bm{\theta}^*$, satisfying:
    \begin{equation}\label{eq12}
        \pi_i^p(\theta_i^{*},\mu) \geq \pi_i^p(\theta_i,\mu), \forall \ \theta_i \in \mathbb{R}, i \in \mathcal{N}
        \vspace{-3pt}
    \end{equation}
    Also, the allocation profile  $\bm{q}^*$ is efficient where $\bm{q}^*$ is defined by $q_i^* =Q(\theta_i^{*},\mu)$.
\end{theorem} 
\section{Strategic Prosumers and Nash Equilibrium} \label{section 4}
In this section, we analyze the oligopoly case where prosumers are price-anticipating. Each prosumer maximizes the following payoff function which is the same as \eqref{eq10} except that now the prosumer realizes that the price is set as a function of all prosumers' actions according to \eqref{eq9}, i.e., $\mu = p(\boldsymbol{\theta})$:
\vspace{-3pt}
\begin{equation}\label{eq13}
\begin{aligned}
    \pi_i^p(\theta_i,\theta_{-i}) = S_i(Q(\theta_i,p(\boldsymbol{\theta})))- p(\boldsymbol{\theta}) Q(\theta_i,p(\boldsymbol{\theta}))
\end{aligned}
\vspace{-3pt}
\end{equation}
Since the prosumer's payoff is a function of the actions of all prosumers, this incentivises the prosumers to strategically adjust their payoff functions. Let $\mathcal{G}$ denote the game defined by the set of prosumers (players) $\mathcal{N}$, their payoffs given by \eqref{eq13} and their action space $\boldsymbol{\Theta}_i = \mathbb R$. Our goal is to demonstrate that the game $\mathcal{G}$ has a Nash equilibrium and that the corresponding market allocation is unique, providing an efficient way to compute it. This can be achieved by showing that at a Nash equilibrium, the resulting allocation is obtained by solving a modified version of the convex optimization problem \eqref{eq7} where the prosumers modify their utility/cost functions $S_i(q_i)$. For notational simplicity, we use slight abuse of notations to refer to $Q(\theta_i,p(\boldsymbol{\theta}))$ or $Q(\theta_i,\mu)$ as $q_i$ and $\pi_i^p(\theta_i,\theta_{-i})$ or $\pi_i^p(\theta_i,\mu)$ as $\pi_i^p$.

The collection of parameters $\boldsymbol{\tilde{\theta}}$ (i.e. bidding profile) constitutes a Nash equilibrium for the game $\mathcal{G}$ if:
\vspace{-2pt}
\begin{equation}\label{eq14}
\pi_i^p(\tilde{\theta_i},\tilde{\theta_{-i}}) \geq \pi_i^p(\theta_i,\tilde{\theta_{-i}}), \forall \theta_i \in \mathbb{R}, \ i \in \mathcal{N}
\vspace{-3pt}
\end{equation}
First, we state some conditions on the prosumers' action spaces in which the existence of Nash equilibrium for the game $\mathcal{G}$ is ruled out:
\vspace{-2pt}
\begin{lemma}\label{lemma1}
If $\tilde{\boldsymbol{\theta}}$ is a Nash equilibrium for the game $\mathcal{G}$, then the following cannot hold: $\tilde{\boldsymbol{\theta}} = \boldsymbol{0}$ or $\forall i \in \mathcal{N}, \sum_{j\neq i}^N \tilde{\theta_j} \geq 0$.
\end{lemma}
The proof is given in Appendix \ref{Plmma1}. It is worth noting from the proof that prosumer $i$ can exert market power if $\sum_{j\neq i}^N \tilde{\theta_j} >0$ since its payoff would increase without a bound. Let $(S_i)_{q_i}$ and $(S_i)_{q_iq_i}$ denote, respectively, the first and second derivatives of $S_i$ with respect to $q_i$. Next, we state a sufficient condition on the prosumers' action spaces for the existence of Nash equilibrium for the game $\mathcal{G}$:
\vspace{-2pt}
\begin{lemma}\label{lemma2}
Assume that $N\geq 2$, and suppose that Assumption \ref{assumption1} holds. Then, $\mathcal{G}$ admits a Nash equilibrium $\boldsymbol{\tilde{\theta}}$ where the following condition is satisfied for all $i \in \mathcal{N}$:
\vspace{-5pt}
\begin{align} \raisetag{8pt}\label{eq15}
-(\sum_{j\neq i}^N \tilde{\theta_j})\Big(\frac{\scalemath{0.9}{Nd_{min}}}{2}\frac{(S_i)_{\tilde{q_i}\tilde{q_i}}(\tilde{q_i})}{(S_i)_{\tilde{q_i}}(\tilde{q_i})}+1\Big) \leq \tilde{\theta_i} \leq -(\sum_{j\neq i}^N \tilde{\theta_j}) - \epsilon
\vspace{-1cm}
\end{align}
where $\tilde{q_i} = Q(\tilde{\theta_i},p(\boldsymbol{\tilde{\theta}}))$ as defined in \eqref{eq6}, $\epsilon>0$ is any infinitesimal constant.
\vspace{-5pt}
\end{lemma}
The proof is given in Appendix \ref{Plmma2}. It is worth noting from the proof that the left inequality of condition \eqref{eq15} represents the interval in which the prosumer's payoff \eqref{eq13} is concave in $\theta_i$ and the right inequality represents the interval in which the payoff is continuous in $\theta_i$---at $\theta_i=-(\sum_{j\neq i}^N \theta_j)$, the market price \eqref{eq9} is zero and the payoff \eqref{eq13} is undefined; hence, $\epsilon>0$ is a technical requirement enforced by the market operator which guarantees the continuity of \eqref{eq13} over a compact subset of $\mathbb{R}$ for $\theta_i$, and ensures a positive market price. We can now state the main result, concluding the uniqueness of Nash equilibrium and characterizing its corresponding market allocation. To prove this result, we construct a convex optimization problem by modifying \eqref{eq7} such that we replace the utility/cost functions $S_i(q_i)$ by modified utility/cost functions $\tilde{S_i}(q_i)$:
\begin{subequations}\label{eq16}
\vspace{-5pt}
\begin{alignat}{2}
    & \underset{\boldsymbol{q}} {\text{maximize}} \ \ \mathcal{\tilde{W}}(\boldsymbol{q})\coloneqq \ \ &&  \sum_{i=1}^N  \tilde{S_i}(q_i), \label{eq16a} \\
    & \text{subject to} &&  \sum_{i=1}^N q_i = 0, \label{eq16b} \\
    &&&  s_{max} \ge -q_i, \ \forall i \in \mathcal{N} \label{eq16c} 
\end{alignat}
\end{subequations}
\vspace{-7pt} where
\begin{gather}\label{eq17} 
\begin{aligned}
\raisetag{27pt}
\tilde{S_i}(q_i) = \begin{cases} \big(1+\frac{q_i}{(N-1)d_{min}}\big) S_i(q_i) - \frac{\int_{d_{min}}^{q_i} S_i(z) \,dz}{(N-1)d_{min}} , q_i {\geq} d_{min} \\[10pt]
\big(1+\frac{q_i}{(N-1)d_{min}}\big) S_i(q_i) + \frac{\int_{q_i}^{d_{min}} S_i(z) \,dz}{(N-1)d_{min}}, q_i {<} d_{min}
\end{cases}
\end{aligned}
\end{gather}
To state the result in Theorem 2, another assumption on $S_i$ is needed which guarantees a unique solution to \eqref{eq16}:
\begin{assumption}\label{assumption2} 
Let $N\geq2$ and $\forall i \in \mathcal{N}$, $q_i,q_i' \in \mathbb{R}$, $d_{min},s_{max} \in \mathbb{R}^+$, and $-s_{max} \leq q_i < q_i'$, $S_i(q_i)$ satisfies the following condition which is stricter than strict concavity:
\small \begin{equation}\label{eq8}
(1 + \frac{q_i'}{\scalemath{0.9}{(N-1)d_{min}}}) \frac{\partial S_i(q_i')}{\partial q_i'} < (1 + \frac{q_i}{\scalemath{0.9}{(N-1)d_{min}}}) \frac{\partial S_i(q_i)}{\partial q_i}
\end{equation}
\end{assumption}
\vspace{-6pt}
\begin{theorem}\label{theorem2}
Assume that $N\geq 2$ and suppose that Assumptions \ref{assumption1} and \ref{assumption2} hold. Let $\boldsymbol{\tilde{q}}$ be an allocation profile corresponding to a Nash equilibrium $\boldsymbol{\tilde{\theta}}$ for the game $\mathcal{G}$, i.e., $\forall i \in \mathcal{N}, \tilde{q_i} = Q(\tilde{\theta_i},p(\tilde{\boldsymbol{\theta}}))$ as defined in \eqref{eq6}. If 
\begin{equation}\label{eq18}
\tilde{q_i} \geq -(N-1)d_{min} - \frac{(S_i)_{\tilde{q_i}}(\tilde{q_i})}{(S_i)_{\tilde{q_i}\tilde{q_i}}(\tilde{q_i})}, \forall i \in \mathcal{N}
\end{equation}
then $\boldsymbol{\tilde{q}}$ is the unique solution to the convex optimization problem \eqref{eq16} and 
 $\boldsymbol{\tilde{\theta}}$ is unique.
\vspace{-4pt}
\end{theorem} 
The proof is given in Appendix \ref{Pthm2}. It is worth noting from the proof that the condition \eqref{eq18} guarantees the concavity of the modified utility/cost function \eqref{eq17} in $q_i$. Theorem 2 provides an efficient way of computing the Nash equilibrium for the game $\mathcal{G}$. Rather than solving $N$ prosumer problems in the action variables $\boldsymbol{\theta}$, we can compute the solution of the optimization problem \eqref{eq16}, providing the market allocation $\boldsymbol{\tilde{q}}$. This, in turn, allows the computation of Nash equilibrium $\boldsymbol{\tilde{\theta}}$ directly using \eqref{eq6}. To understand the rationale for constructing the optimization problem \eqref{eq16}, first note that it is similar to \eqref{eq7} except the objective function, where the utility/cost functions of the prosumers are modified. Therefore, \eqref{eq16a} represents the maximization of a welfare at the Nash equilibrium which is not the true welfare maximized in \eqref{eq7a}. This means that at the Nash equilibrium, the true utility/cost functions $S_i(q_i)$ are strategically misrepresented by the prosumers such that they declare untruthful utility/cost functions $\tilde{S_i}(q_i)$ to maximize their profits. 
\vspace{-0.20cm}
\section{Case Study} \label{section 5}
In this section, our goal is to examine the welfare loss due to the strategic behavior of $N$ prosumers when the market's supply capacity or inelastic demand are varied. To achieve this, we compute the social welfare and contrast its behavior under two scenarios: first with the optimal allocation resulting from the perfect competition of the prosumers, given by the program \eqref{eq7}, and second with the optimal allocation resulting from the prosumers' strategic interaction, given by the program \eqref{eq16}. In both cases, we calculate the true social welfare defined in \eqref{eq7a}. We vary the market's supply capacity or inelastic demand by changing the supply $s_{max}$ or demand $d_{min}$ parameters in the proposed bidding mechanism \eqref{eq6}, respectively, while fixing the other parameters. We select the range in which we vary these parameters such that the welfare of the perfect competition plateaus. We consider the following example of a strictly concave, strictly increasing utility/cost function $S_i(q_i)$ for each prosumer $i \in \mathcal{N}$:
\begin{equation} \label{eq19}
S_i(q_i)= e^{\Large \frac{-\beta_i}{5}} - e^{\Large \frac{- \beta_i  q_i}{5 d_{min}}} 
\end{equation}
Using this example function, we can compute the true welfare defined in \eqref{eq7a}. Also, to calculate the ``modified'' welfare \eqref{eq16a}, we write $\tilde{S_i}(q_i)$, defined in \eqref{eq17}, as follows:
\begin{gather}
\begin{aligned} \label{eq20} \raisetag{20pt}
\tilde{S_i}(q_i) =& (1+\scaleto{\frac{q_i}{(N-1) d_{min}}}{17pt})\cdot  ( e^{\Large \frac{-\beta_i}{5}} - e^{\Large \frac{- \beta_i  q_i}{5 d_{min}}} ) - \scaleto{\frac{1}{(N-1)d_{min}}}{17pt} \\& \cdot  \Big( e^{\Large \frac{-\beta_i}{5}}(q_i - d_{min}) + \scaleto{\frac{5 d_{min}}{\beta_i}}{17pt} (e^{\Large \frac{-\beta_i q_i}{5 d_{min}}} - e^{\Large \frac{-\beta_i}{5}}) \Big)
\end{aligned}
\end{gather}
\normalsize
To compare the values of the two resulting welfares (welfare at the competitive equilibrium and welfare at the Nash equilibrium) when the market's supply/demand parameters are changed, we solve both programs \eqref{eq7} and \eqref{eq16} several times, first varying the total supply capacity (i.e. $s_{max}$ for all $i \in \mathcal{N}$) and second changing the total inelastic demand (i.e. $d_{min}$ for all $i \in \mathcal{N}$). In both simulations and using an ad-hoc technique, we investigate two cases. In the first case, we make sure that the conditions in Lemma \ref{lemma1} and Lemma \ref{lemma2} are satisfied. That is, we check in each simulation if $\forall i \in \mathcal{N}$, $\sum_{j\neq i}^N \tilde{\theta_j} < 0$ and \eqref{eq15} are satisfied by tuning the parameters $\beta_i$, $d_{min}$, and $s_{max}$. In the second case, we carry out other simulations such that we allow the left inequality of \eqref{eq15} to not be satisfied; our goal is to observe whether the welfare loss would be the same as in the first case where \eqref{eq15} is always satisfied---recall that the left inequality of \eqref{eq15} is a sufficient condition for the existence of Nash equilibrium since it guarantees concavity of each prosumer's payoff in $\theta_i$. Also, recall that both \eqref{eq18} and Assumption \ref{assumption2} constitute the sufficient conditions for existence and uniqueness of the market allocation at Nash equilibrium, since \eqref{eq18} guarantees concavity of each prosumer's modified utility/cost function in $q_i$ while Assumption \ref{assumption2} guarantees strict concavity of the objective function \eqref{eq16a}. It is worth noting that the set $\mathcal{B}$, constituting all possible $\theta_i$'s defined by the inequality in $\theta_i$ that is obtained from substituting $Q(\theta_i,p(\boldsymbol{\theta}))$ in \eqref{eq18}, is contained within the set $\mathcal{A}$ defined by the left inequality of \eqref{eq15}---see Appendix \ref{Pthm2} for more details. Given our example function \eqref{eq19}, the condition \eqref{eq18} yields:
\begin{equation} \label{eq21}
\tilde{q_i} \geq \frac{5d_{min}}{\beta_i} - d_{min}(N-1), i \in \mathcal{N}
\vspace{-3pt}
\end{equation}
\eqref{eq21} indicates that to guarantee existence of Nash equilibrium, the optimal allocation for each prosumer $i$ must be above a certain value which depends on $\beta_i$, $d_{min}$, and $N$. In all the simulations, we fix the number of prosumers $N$ to 11. By fixing $d_{min}>0$, the right-hand side of \eqref{eq21} decreases as $\beta_i$ increases. Similarly by fixing $\beta_i>0.5$, it decreases as $d_{min}$ increases. Therefore, the minimum optimal allocation that is sufficient for existence of Nash equilibrium moves further left in the real line as we increase $d_{min}$ and/or $\beta_i$. When the absolute value of this minimum value is less than the maximum supply capacity $s_{max}$ for all prosumers, existence of Nash equilibrium is guaranteed.

\begin{figure}[h]%
    \centering
    \vspace{0.21cm}
    {{\includegraphics[width=8.1cm]{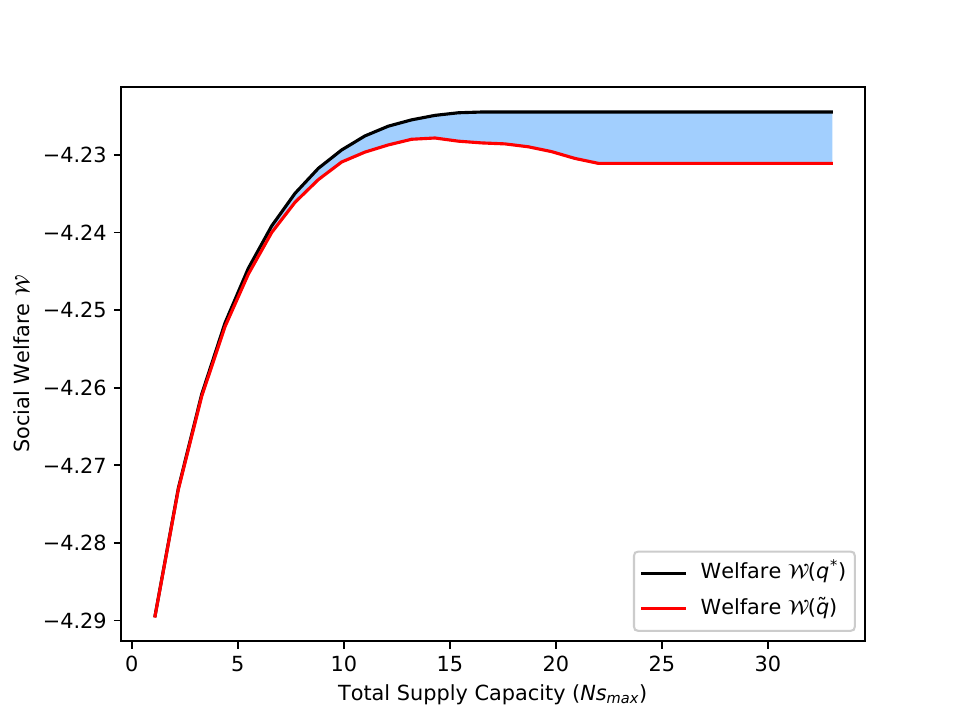} }}%
    \enspace
    {{\includegraphics[width=8.1cm]{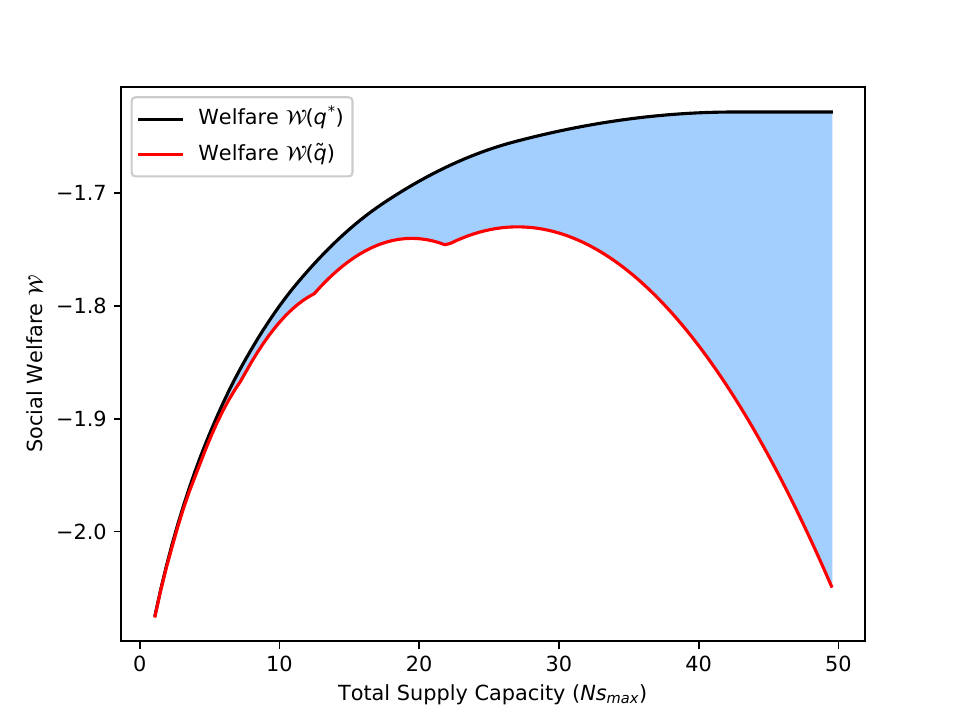}}}%
    \vspace{-2pt}
    \caption{Welfare loss with increasing supply capacity and fixed inelastic demand. (Top) Nash equilibrium exists and the welfare loss does not grow unbounded. (Bottom) Nash equilibrium may not exist and the welfare loss grows unbounded.}%
    \label{figure1}%
    \vspace{-0.7cm}
\end{figure}

\begin{figure}[h]%
    \centering
    \vspace{0.21cm}
    {{\includegraphics[width=8.1cm]{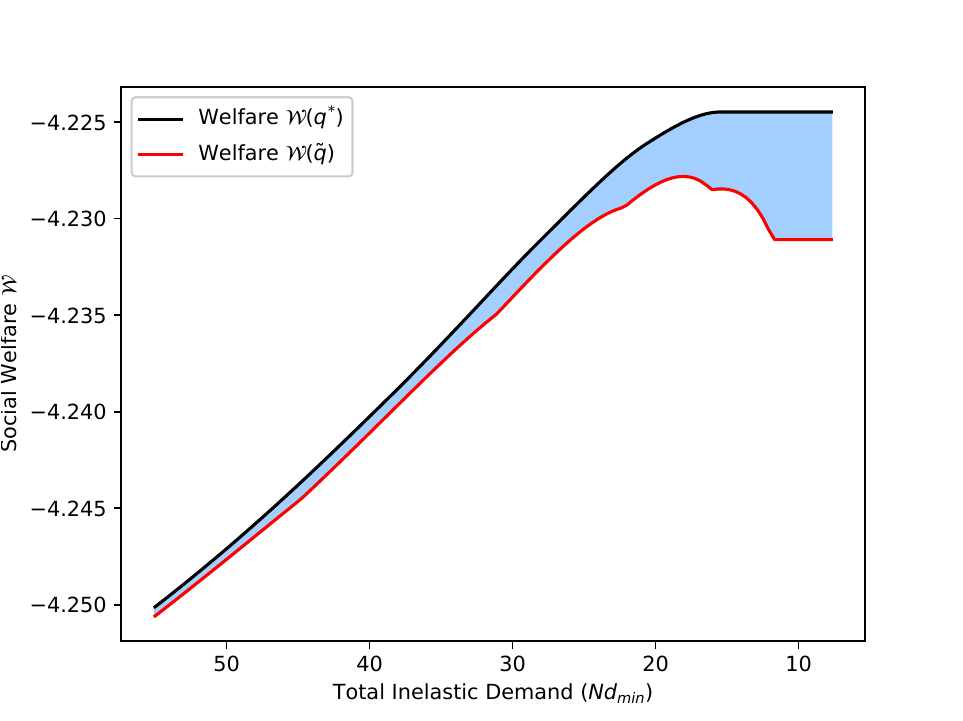} }}%
    \enspace
    {{\includegraphics[width=8.1cm]{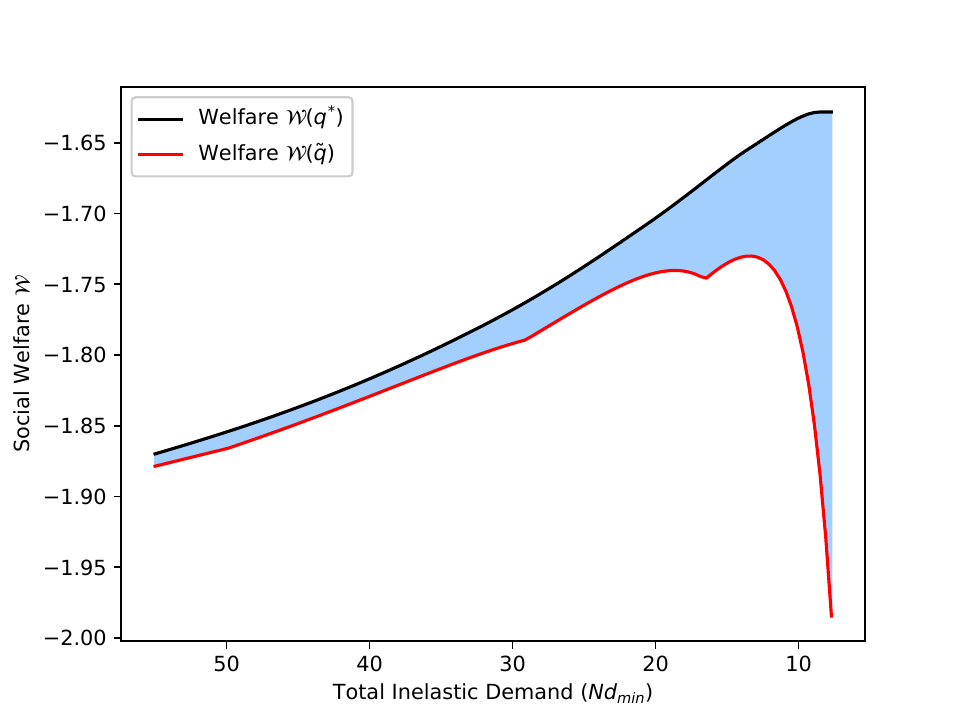} }}%
    \vspace{-2pt}
    \caption{Welfare loss with decreasing inelastic demand and fixed supply capacity. (Top) Nash equilibrium exists and the welfare loss does not grow unbounded. (Bottom) Nash equilibrium may not exist and the welfare loss grows unbounded.}%
    \label{figure2}%
    \vspace{-0.7cm}
\end{figure}

First, we compare the two resulting welfares and observe the gap between them when the prosumers' supply capacities $s_{max}$ are varied while fixing their inelastic demands $d_{min}$. On the top of Fig. \ref{figure1}, $s_{max}$ is increased gradually from 0.1 to 3 while $d_{min}=4$ and $\beta_i = \left\{1.9+0.1i \;\middle\vert\; i=1,...,N\right\}$ are fixed. As mentioned earlier, the values of the latter two are selected using an ad-hoc technique to guarantee existence of Nash equilibrium by making sure that the resulting optimal allocations of the prosumers always satisfy \eqref{eq21}. The figure shows that if a Nash equilibrium exists, the welfare loss does not grow unbounded when the total supply capacity is increased. Similarly, Fig. \ref{figure1} on the bottom shows the welfare gap when $s_{max}$ is increased from 0.1 to 4.5 while $d_{min}=1$ and $\beta_i = \left\{0.5+0.1i \;\middle\vert\; i=1,...,N\right\}$ are fixed. In contrast to the previous simulations, in this case, the values of $d_{min}$ and $\beta_i$ are selected such that the resulting optimal allocations of the prosumers do not all necessarily satisfy \eqref{eq21} when their supply capacities exceed a certain threshold. Consequently, a Nash equilibrium may not exist. The figure shows that the welfare loss grows unbounded when the total supply capacity is increased. In such simulations, the optimal allocations of prosumers 1 and 2 do not satisfy \eqref{eq21} when the total supply capacity approximately exceeds 18.5 and 31.5, respectively.
Second, we examine the welfare loss between the two resulting welfares when the prosumers' inelastic demands $d_{min}$ are varied while having their supply capacities $s_{max}$ fixed. On the top of Fig. \ref{figure2}, $s_{max}=0.7$ and $\beta_i = \left\{1.9+0.1i \;\middle\vert\; i=1,...,N\right\}$ are fixed while $d_{min}$ is decreased gradually from 5 to 0.7. The values of the latter two are selected using an ad-hoc technique so that the condition \eqref{eq21} for the existence of Nash equilibrium is satisfied for all prosumers. The figure shows that when the total inelastic demand is decreased, the welfare loss does not grow unbounded if a Nash equilibrium exists. Similarly, Fig. \ref{figure2} on the bottom shows the welfare gap when $s_{max}=3$ and $\beta_i = \left\{0.5+0.1i \;\middle\vert\; i=1,...,N\right\}$ are fixed while $d_{min}$ is decreased from 5 to 0.7. The values of the latter two are selected such that, in this case, the resulting optimal allocations of the prosumers do not all necessarily satisfy \eqref{eq21} when their inelastic demands are below a certain threshold. Thus, a Nash equilibrium may not exist. The figure shows that the welfare loss grows unbounded when the total inelastic demand is decreased. In these simulations, the optimal allocations of prosumers 1 and 2 do not satisfy \eqref{eq21} when the total inelastic demand drops approximately below 20 and 11.5, respectively.
\newline
\vspace{-0.40cm}
\section{Conclusion and Future Directions} \label{section 6}
In this paper, a scalar-parameterized bidding mechanism has been proposed for the prosumers in a uniform-price peer-to-peer market. A competitive equilibrium and the associated efficient allocation have been established. When certain conditions on the action spaces of the prosumers are satisfied, we have shown that a unique Nash equilibrium exists. In addition, we have provided an efficient way to compute the market allocation at the Nash equilibrium, and in turn, the Nash equilibrium itself. Finally, a case study was given where we have shown that the welfare gap between the welfare at the competitive equilibrium and the welfare at the Nash equilibrium is bounded when the market supply or inelastic demand are varied. On the contrary when the existence of Nash equilibrium is not guaranteed, the welfare loss grows unbounded as the market supply is increased or the market inelastic demand is decreased. A future research direction would be to characterize a bound for the welfare loss. Also given that the proposed mechanism does not allow the prosumers to choose their preferred quantity of supply and demand separately, a future research direction would be to develop a mechanism which captures the dual nature of the prosumers. 
\vspace{-6pt}
\bibliographystyle{IEEEtran}
\bibliography{content/Bibliography}

\begin{thebibliography}{10}
\providecommand{\url}[1]{#1}
\csname url@samestyle\endcsname
\providecommand{\newblock}{\relax}
\providecommand{\bibinfo}[2]{#2}
\providecommand{\BIBentrySTDinterwordspacing}{\spaceskip=0pt\relax}
\providecommand{\BIBentryALTinterwordstretchfactor}{4}
\providecommand{\BIBentryALTinterwordspacing}{\spaceskip=\fontdimen2\font plus
\BIBentryALTinterwordstretchfactor\fontdimen3\font minus
  \fontdimen4\font\relax}
\providecommand{\BIBforeignlanguage}[2]{{%
\expandafter\ifx\csname l@#1\endcsname\relax
\typeout{** WARNING: IEEEtran.bst: No hyphenation pattern has been}%
\typeout{** loaded for the language `#1'. Using the pattern for}%
\typeout{** the default language instead.}%
\else
\language=\csname l@#1\endcsname
\fi
#2}}
\providecommand{\BIBdecl}{\relax}
\BIBdecl

\bibitem{kelly1998rate}
F.~P. Kelly, A.~K. Maulloo, and D.~K.~H. Tan, ``Rate control for communication
  networks: shadow prices, proportional fairness and stability,'' \emph{Journal
  of the Operational Research society}, vol.~49, pp. 237--252, 1998.

\bibitem{johari2004efficiency}
R.~Johari and J.~N. Tsitsiklis, ``Efficiency loss in a network resource
  allocation game,'' \emph{Mathematics of Operations Research}, vol.~29, no.~3,
  pp. 407--435, 2004.

\bibitem{vickrey1961counterspeculation}
W.~Vickrey, ``Counterspeculation, auctions, and competitive sealed tenders,''
  \emph{The Journal of finance}, vol.~16, no.~1, pp. 8--37, 1961.

\bibitem{clarke1971multipart}
E.~H. Clarke, ``Multipart pricing of public goods,'' \emph{Public choice}, pp.
  17--33, 1971.

\bibitem{groves1973incentives}
T.~Groves, ``Incentives in teams,'' \emph{Econometrica: Journal of the
  Econometric Society}, pp. 617--631, 1973.

\bibitem{maheswaran2003nash}
R.~T. Maheswaran and T.~Ba{\c{s}}ar, ``Nash equilibrium and decentralized
  negotiation in auctioning divisible resources,'' \emph{Group Decision and
  Negotiation}, vol.~12, no.~5, pp. 361--395, 2003.

\bibitem{maheswaran2006efficient}
R.~Maheswaran and T.~Ba{\c{s}}ar, ``Efficient signal proportional allocation
  (espa) mechanisms: Decentralized social welfare maximization for divisible
  resources,'' \emph{IEEE Journal on Selected Areas in Communications},
  vol.~24, no.~5, pp. 1000--1009, 2006.

\bibitem{johari2011parameterized}
R.~Johari and J.~N. Tsitsiklis, ``Parameterized supply function bidding:
  Equilibrium and efficiency,'' \emph{Operations research}, vol.~59, no.~5, pp.
  1079--1089, 2011.

\bibitem{klemperer1989supply}
P.~D. Klemperer and M.~A. Meyer, ``Supply function equilibria in oligopoly
  under uncertainty,'' \emph{Econometrica: Journal of the Econometric Society},
  pp. 1243--1277, 1989.

\bibitem{lin2019structural}
W.~Lin and E.~Bitar, ``A structural characterization of market power in
  electric power networks,'' \emph{IEEE Transactions on Network Science and
  Engineering}, vol.~7, no.~3, pp. 987--1006, 2019.

\bibitem{ndrio2020scalar}
M.~Ndrio, K.~Alshehri, and S.~Bose, ``A scalar-parameterized mechanism for
  two-sided markets,'' \emph{IFAC-PapersOnLine}, vol.~53, no.~2, pp.
  16\,952--16\,957, 2020.

\bibitem{parag2016electricity}
Y.~Parag and B.~K. Sovacool, ``Electricity market design for the prosumer
  era,'' \emph{Nature energy}, vol.~1, no.~4, pp. 1--6, 2016.

\bibitem{saad2012game}
W.~Saad, Z.~Han, H.~V. Poor, and T.~Ba{\c{s}}ar, ``Game-theoretic methods for
  the smart grid: An overview of microgrid systems, demand-side management, and
  smart grid communications,'' \emph{IEEE Signal Processing Magazine}, vol.~29,
  no.~5, pp. 86--105, 2012.

\bibitem{bacsar1998dynamic}
T.~Ba{\c{s}}ar and G.~J. Olsder, \emph{Dynamic Noncooperative Game
  Theory}.\hskip 1em plus 0.5em minus 0.4em\relax SIAM, 1998.

\bibitem{alshehri2020quantifying}
K.~Alshehri, M.~Ndrio, S.~Bose, and T.~Ba{\c{s}}ar, ``Quantifying market
  efficiency impacts of aggregated distributed energy resources,'' \emph{IEEE
  Transactions on Power Systems}, vol.~35, no.~5, pp. 4067--4077, 2020.

\bibitem{gharesifard2015price}
B.~Gharesifard, T.~Ba{\c{s}}ar, and A.~D. Dom{\'\i}nguez-Garc{\'\i}a,
  ``Price-based coordinated aggregation of networked distributed energy
  resources,'' \emph{IEEE Transactions on Automatic Control}, vol.~61, no.~10,
  pp. 2936--2946, 2015.

\bibitem{xu2015demand}
Y.~Xu, N.~Li, and S.~H. Low, ``Demand response with capacity constrained supply
  function bidding,'' \emph{IEEE Transactions on Power Systems}, vol.~31,
  no.~2, pp. 1377--1394, 2015.

\bibitem{tsybina2022effect}
E.~Tsybina, J.~Burkett, and S.~Grijalva, ``The effect of prosumer duality on
  power market: evidence from the cournot model,'' \emph{IEEE Transactions on
  Power Systems}, vol.~38, no.~1, pp. 692--701, 2022.

\end{thebibliography}
\useRomanappendicesfalse
\begin{appendices}
\vspace{-5pt}
\renewcommand{\thesectiondis}[2]{\Alph{section}:}
\section{Proof of Theorem 1} \label{Pthrm1}
We want to show that the first fundamental theorem of welfare economics holds for the proposed mechanism. This can be achieved by a standard approach in which the equilibrium conditions of \eqref{eq11} together with $ \mu>0$ given by \eqref{eq9} is shown to be equivalent to the optimality conditions of the program \eqref{eq7} with a one-to-one correspondence. We first derive the optimality conditions for the prosumer's problem, and then derive the optimality conditions for the market operator's problem. Given that $q_i$ is given by \eqref{eq6} and since the prosumer is constrained by a maximum supply capacity $s_{max}$ (i.e. $-s_{max}\le q_i$) along with the demand/supply balance constraint $\sum_{i=1}^N q_i = 0$, then the action variable $\theta_i$ belongs to a compact, convex subset of $\mathbb R$. Given a price $\mu>0$, the second derivative of the prosumer's payoff \eqref{eq11} with respect to the action variable $\theta_i$ is given by:
\begin{equation}\label{eq22}
\frac{\partial^2 \pi_i^p}{\partial \theta_i^2} = \frac{\partial^2 S_i(Q(\theta_i,\mu))}{\partial q_i^2}\Big(\frac{1}{\mu}\Big)^2
\end{equation}
Since Assumption \ref{assumption1} implies that the second derivative of $S_i$ is non-positive, then \eqref{eq22} is non-positive. Hence, the prosumer's payoff \eqref{eq11} is concave in $\theta_i$. Also, \eqref{eq11} is continuous by Assumption \ref{assumption1}. As a result and since the Slater's constraint qualification (strict feasibility) holds (i.e. there exists a $\theta_i$ such that $\theta_i{>}\mu(-s_{max}-d_{min})$), then the Karush-Kuhn-Tucker (KKT) conditions are necessary and sufficient for optimality for \eqref{eq11}. Associate the Lagrange multiplier $\lambda$ with the inequality constraint for $\theta_i$ obtained from $-s_{max}\le q_i$ and given by $\theta_i{\ge}\mu(-s_{max}-d_{min})$. Thus, the Lagrangian function becomes:
\vspace{-4pt}
\begin{equation} \label{eq23}
L(\theta_i,\lambda) = S_i(q_i) - q_i  \mu + \lambda  (\theta_i - \mu (-s_{max}-d_{min}))
\vspace{-4pt}
\end{equation}
Let $\theta_i^*$ denote an optimal action. Analyzing the KKT conditions shows that $\theta_i^*$ must satisfy:
\vspace{-2pt}
\begin{subequations}\label{eq24}
\begin{equation}\label{eq24a} 
\frac{\partial S_i(Q(\theta_i^*,\mu))}{\partial q_i} = \mu,\ \text{if} \  \theta_i^* > \mu(-s_{max}-d_{min})
\end{equation}
\vspace{-7pt}
\begin{equation}\label{eq24b} 
\frac{\partial S_i(Q(\theta_i^*,\mu))}{\partial q_i} < \mu,\ \text{if} \ \theta_i^* = \mu(-s_{max}-d_{min})
\vspace{-6pt}
\end{equation}
\end{subequations}
Next, we analyze the convex optimization problem \eqref{eq7} that is solved by the market operator. The constraint \eqref{eq7c} along with the demand/supply balance constraint \eqref{eq7b} make the feasible solution set compact and convex. Therefore, the objective function is strictly concave and continuous in $q_i$, by Assumption 1, over a compact convex set. And since the Slater's constraint qualification holds (i.e. there exists a $q_i$ such that $-s_{max}{<} q_i, \forall i \in \mathcal{N}$), then the KKT conditions are necessary and sufficient for the solution of \eqref{eq7} to be optimal and unique. Associate the Lagrange multipliers $\eta$ with the equality constraint \eqref{eq7b} and $\lambda_i, i \in \mathcal{N}$ with the inequality constraints \eqref{eq7c}. Thus, the Lagrangian function becomes:
\begin{equation} \label{eq25}
L(\boldsymbol{q},\eta, \boldsymbol{\lambda}) = \sum_{i=1}^N S_i(q_i) - \eta \Big(\sum_{i=1}^N q_i\Big) + \sum_{i=1}^N \lambda_i (q_i + s_{max})
\end{equation}
Analyzing the KKT conditions gives the conditions for the unique optimal solution $\boldsymbol{q}^*$ and $\eta^* \geq0$. The vector $\boldsymbol{q}^*$ must satisfy:
\vspace{-6pt}
\begin{subequations}\label{eq26}
\begin{equation}\label{eq26a}
\frac{\partial S_i(q_i^{*})}{\partial q_i} = \eta^*,\ \text{if} \ q_i^{*} > -s_{max}
\vspace{-5pt}
\end{equation}
\begin{equation}\label{eq26b}
\frac{\partial S_i(q_i^{*})}{\partial q_i} < \eta^*,\ \text{if} \  q_i^{*} = -s_{max}
\end{equation}
\end{subequations}
\vspace{-0.1cm}
Also, the demand/supply balance constraint \eqref{eq7b} must hold for $q_i^*$: 
\vspace{-9pt}
\begin{equation}\label{eq27}
\sum_{i=1}^N q_i^* = 0
\vspace{-0.1cm}
\end{equation}
It is easy to show that if we let, by \eqref{eq6}, $\theta_i = \eta^*(q_i^*-d_{min})$, then $(\boldsymbol{q}^*,\eta^*)$ satisfying \eqref{eq26} is equivalent to $(\boldsymbol{\theta},\eta^*)$ satisfying \eqref{eq24}. Also, \eqref{eq27} implies that $\eta^*=\mu$. Therefore, $(\boldsymbol{\theta},\mu)$ constitute a competitive equilibrium. Similarly it is easy to verify that if we let, by \eqref{eq6}, $q_i = Q(\theta_i^*,\mu)=d_{min}+\frac{\theta_i^*}{\mu}$, then $(\boldsymbol{\theta}^*,\mu)$ satisfying \eqref{eq24} and \eqref{eq9} is equivalent to  $\boldsymbol{q}$ satisfying \eqref{eq26}. Therefore, $\boldsymbol{q}$ is an efficient allocation. Finally, uniqueness of the competitive equilibrium $(\boldsymbol{\theta}^*,\mu)$ follows from its one-to-one correspondence with the unique optimal allocation $\boldsymbol{q}^*$.
\vspace{-0.2cm}
\section{Proof of Lemma 1} \label{Plmma1}
The first derivative of the prosumer payoff \eqref{eq13} with respect to the action variable $\theta_i$ is given by:
\small \begin{equation}\label{eq28}
    \frac{\partial \pi_i^p(\theta_i,\theta_{-i})}{\partial \theta_i} =\frac{\partial S_i(Q(\theta_i,p(\boldsymbol{\theta})))}{\partial q_i}  \frac{\partial Q(\theta_i,p(\boldsymbol{\theta}))}{\partial \theta_i} {+} \frac{1}{N} {-} 1
\end{equation}
\normalsize
where:
\vspace{-10pt}
\begin{equation}\label{eq29}
    \frac{\partial Q(\theta_i,p(\boldsymbol{\theta}))}{\partial \theta_i} =\frac{-(\sum_{j\neq i}^N \theta_j)Nd_{min} }{(\sum_{j=1}^N \theta_j)^2}
\vspace{-5pt}
\end{equation}
Recall the assumption $\sum_{j= 1}^N \theta_j{\leq}0$ from \eqref{eq9}. Let $\sum_{j= 1}^N \theta_j{<}0$. If $\sum_{j\neq i}^N \theta_j>0$, then \eqref{eq29} becomes strictly negative. Hence, \eqref{eq28} is strictly negative when Assumption \ref{assumption1} is satisfied. In other words, as $\theta_i$ is decreased (i.e. supply is increased), the payoff \eqref{eq13} increases without bound. Therefore, a Nash equilibrium cannot exist in this case. To investigate the two cases: $\boldsymbol{\theta}=\boldsymbol{0}$ and/or $\sum_{j\neq i}^N \theta_j=0$ and $\theta_i\ne0$, we rewrite the payoff \eqref{eq13}:
\vspace{-10pt}
\begin{equation}\label{eq30} \begin{split}
    \pi_i^p(\theta_i,\theta_{-i}) = S_i(d_{min}(1-\frac{\theta_i}{\sum_{j=1}^N \theta_j}N))+ \\ \frac{\sum_{j=1}^N \theta_j}{N} (1-\frac{\theta_i}{\sum_{j=1}^N \theta_j}N)
\vspace{-5pt}
\end{split}
\end{equation}
Both terms in \eqref{eq30} can be positive or negative depending on the sign of $\theta_i$. If $\theta_i=0$, then the first term is zero and the second term is either zero or negative depending on whether $\sum_{j=1}^N \theta_j$ is zero or negative, respectively. Therefore, a Nash equilibrium cannot exist in the first case (i.e. $\boldsymbol{\theta}=\boldsymbol{0}$) since the payoff \eqref{eq30} can be increased from zero to a positive value by deviating from $\theta_i=0$. As a result, $\sum_{j=1}^N \theta_j{=}0$ is not possible. In the second case (i.e. $\sum_{j\neq i}^N \theta_j=0$ and $\theta_i\ne0$), the payoff \eqref{eq30} can be increased from a negative value to zero by deviating from $\theta_i>0$ or $\theta_i<0$ to $\theta_i=0$. Therefore, a Nash equilibrium cannot exist in the second case.
\vspace{-0.1cm}
\section{Proof of Lemma 2}\label{Plmma2}
If we let $N=2$ in the first derivative of the prosumer's payoff \eqref{eq13} with respect to $\theta_i$, following the same steps in the proof of Lemma 1, then the first derivative can be negative or positive. Hence, the payoff \eqref{eq13} does not increase without bound as $\theta_i$ is decreased. Therefore, a Nash equilibrium may exist if $N=2$. Next, we want to show that the objective function for the prosumer is continuous and concave over a compact, convex set. Let $Q(\theta_i,p(\boldsymbol{\theta})) = q_i$, i.e., replace $p$ in \eqref{eq6} by $p(\boldsymbol{\theta})$ from \eqref{eq9}. To show that the prosumer's payoff function \eqref{eq13} is concave in $\theta_i$, we examine the condition under which the second derivative of \eqref{eq13} is non-positive. The second derivative of \eqref{eq13} with respect to the action variable $\theta_i$ is given by:
\begin{equation}\label{eq31}
\begin{split}
    \frac{\partial^2 \pi_i^p}{\partial \theta_i^2} =\frac{\partial^2 S_i(Q(\theta_i,p(\boldsymbol{\theta})))}{\partial q_i^2} \Big(\frac{\partial Q(\theta_i,p(\boldsymbol{\theta}))}{\partial \theta_i}\Big)^2 +\\ \frac{\partial S_i(Q(\theta_i,p(\boldsymbol{\theta})))}{\partial q_i} \Big(\frac{-2}{\sum_{j=1}^N \theta_j}\frac{\partial Q(\theta_i,p(\boldsymbol{\theta}))}{\partial \theta_i}\Big)
\end{split}
\end{equation}
By Assumption 1, the first term in \eqref{eq31} is non-positive. And by Assumptions 1 and $\sum_{j=1}^N \theta_j{<}0$, the second term in \eqref{eq31} is non-negative. Making \eqref{eq31} less than or equal to zero yields the following condition:
\vspace{-10pt}
\begin{equation}\label{eq32}
    \theta_i \geq -(\sum_{j\neq i}^N \theta_j)\Big(\frac{Nd_{min}}{2}\frac{\frac{\partial^2 S_i(Q(\theta_i,p(\boldsymbol{\theta})))}{\partial q_i^2}}{\frac{\partial S_i(Q(\theta_i,p(\boldsymbol{\theta})))}{\partial q_i}}+1\Big)
\vspace{-5pt}
\end{equation}
Define the set $\mathcal{A}$ by all possible $\theta_i$ in which \eqref{eq32} is satisfied. If $\theta_i \in \mathcal{A}$, then \eqref{eq31} is non-positive and hence the payoff function \eqref{eq13} is concave in $\theta_i$. Furthermore, \eqref{eq13} is continuous in $\theta_i$ where $\theta_i < -\sum_{j\neq i}^N \theta_j$ (the region derived in Lemma 1 in which Nash equilibrium may exist). When $\theta_i = -\sum_{j\neq i}^N \theta_j$, the price \eqref{eq9} is zero and \eqref{eq13} is undefined. Therefore, the prosumer's payoff \eqref{eq13} is continuous and concave over a compact, convex subset of $\mathbb R$ which is defined by the intersection of \eqref{eq32} and $\theta_i \leq -\sum_{j\neq i}^N \theta_j - \epsilon$ where $\epsilon$ is any infinitesimal positive constant. To conclude, the above conditions constitute sufficient conditions for the existence of Nash equilibrium for the game $\mathcal{G}$.
\vspace{-0.1cm}
\renewcommand{\thesectiondis}[2]{\Alph{section}:}
\section{Proof of Theorem 2} 
\label{Pthm2}
To establish the uniqueness of market allocation $\boldsymbol{\tilde{q}}$ at a Nash equilibrium $\boldsymbol{\tilde{\theta}}$, we show that the players' equilibrium conditions of \eqref{eq13} with $p(\boldsymbol{\theta})>0$ are equivalent to the optimality conditions of \eqref{eq16} with a one-to-one correspondence. The proof begins with deriving the sufficient conditions for existence of a Nash equilibrium. Then, we establish the existence and uniqueness of the market allocation at a Nash equilibrium. Next, we derive the necessary and sufficient KKT optimality conditions of \eqref{eq16}. Finally, we show the one-to-one correspondence of the optimality conditions of \eqref{eq16} to the equilibrium conditions of \eqref{eq13} which concludes the uniqueness of a Nash equilibrium, completing the proof.

\textbf{Sufficient Conditions for Nash Equilibria:}
\vspace{0.1cm}

We showed, in the proof of Lemma 2, the conditions under which \eqref{eq13} is concave and continuous over a compact, convex set which result in a sufficient condition for the existence of Nash equilibria. Since the Slater's constraint qualification holds (i.e. there exists a $\theta_i$ such that $\theta_i{>}p(\boldsymbol{\theta})(-s_{max}-d_{min})$), then the KKT optimality conditions of \eqref{eq13} are necessary and sufficient. Hereafter, we derive these conditions. Associate the Lagrange multiplier $\lambda$ with the inequality constraint for $\theta_i$ obtained from $-s_{max}{\le} q_i$ and given by $\theta_i {\ge} p(\boldsymbol{\theta})(-s_{max}-d_{min})$. Given $p(\boldsymbol{\theta})$ as defined in \eqref{eq9}, the Lagrangian function becomes:\\
\vspace{-5pt}
\small \begin{equation} \label{eq33}
L(\theta_i,\lambda) =  S_i(q_i) - q_i  p(\boldsymbol{\theta}) + \lambda (\theta_i - p(\boldsymbol{\theta}) (-s_{max}-d_{min}))
\end{equation}
\normalsize
Let $\tilde{\theta_i}$ denote a Nash equilibrium. We let, by \eqref{eq6}, $\tilde{\theta_i} = p(\boldsymbol{\tilde{\theta}})(Q(\tilde{\theta_i},p(\boldsymbol{\tilde{\theta}}))-d_{min})$ and invoke $\theta_i < -\sum_{j\neq i}^N \theta_j$ to analyze the KKT conditions. This shows that $\tilde{\theta_i}$ must satisfy:
\vspace{-5pt}
\begin{subequations}\label{eq34}
\begin{equation}\label{eq34a}
\begin{split}
\frac{\partial S_i(Q(\tilde{\theta_i},p(\boldsymbol{\tilde{\theta}})))}{\partial q_i}  \Big(1 + \frac{Q(\tilde{\theta_i},p(\boldsymbol{\tilde{\theta}}))}{(N-1)d_{min}}\Big)  =  p(\boldsymbol{\tilde{\theta}}), \\ \text{if} \ \tilde{\theta_i}>p(\boldsymbol{\tilde{\theta}})(-s_{max}-d_{min}))
\end{split}
\end{equation}
\begin{equation}\label{eq34b}
\begin{split}
\frac{\partial S_i(Q(\tilde{\theta_i},p(\boldsymbol{\tilde{\theta}})))}{\partial q_i}  \Big(1 + \frac{Q(\tilde{\theta_i},p(\boldsymbol{\tilde{\theta}}))}{(N-1)d_{min}}\Big)  <  p(\boldsymbol{\tilde{\theta}}), \\ \text{if} \ \tilde{\theta_i}=p(\boldsymbol{\tilde{\theta}})(-s_{max}-d_{min}))
\end{split}
\end{equation}
\end{subequations}

\textbf{Existence and Uniqueness of the Market Allocation:}
\vspace{0.1cm}

We want to show that the objective function defined in \eqref{eq16a} is continuous and strictly concave over a compact, convex set which implies the existence and uniqueness of the market allocation (i.e. optimal solution to \eqref{eq16}). Let $Q(\theta_i,p(\boldsymbol{\theta})) = q_i$ (i.e. replace $p$ in \eqref{eq6} by $p(\boldsymbol{\theta})$ from \eqref{eq9}). The feasible set is a compact, convex subset of $\mathbb R$ since the prosumer is constrained by a maximum supply capacity $s_{max}$ (i.e. $-s_{max}\le q_i$) and the demand/supply balance constraint $\sum_{i=1}^N q_i = 0$. Next, the second derivative of $\tilde{S_i}(q_i)$ with respect to $q_i$ (hence the objective function \eqref{eq16a}) is given by:
\vspace{-10pt}
\begin{equation}\label{eq35}
\begin{split}
\frac{\partial^2 \tilde{S_i}(q_i)}{\partial q_i^2} = \frac{\partial^2 S_i(q_i)}{\partial q_i^2} \Big(1+\frac{q_i}{(N-1)d_{min}} \Big) + \\ \frac{\partial S_i(q_i)}{\partial q_i} \frac{1}{(N-1)d_{min}}
\end{split}
\vspace{-10pt}
\end{equation}
\normalsize
by Assumption 1, the objective function \eqref{eq16a} is concave in $q_i$ when:
\vspace{-10pt}
\begin{equation}\label{eq36}
q_i \geq -\Bigg( (N-1)d_{min} + \frac{\frac{\partial S_i(q_i)}{\partial q_i}}{\frac{\partial^2 S_i(q_i)}{\partial q_i^2}} \Bigg)
\vspace{-5pt}
\end{equation}
Substitute $Q(\theta_i,p(\boldsymbol{\theta})) = q_i$ in \eqref{eq36} to find the equivalent inequality for $\theta_i$. Let the set $\mathcal{B}$ be defined by all possible $\theta_i$'s in which the equivalent inequality for $\theta_i$ is satisfied. Then, $\mathcal{B} \subset \mathcal{A}$ where $\mathcal{A}$ is the set defined in Lemma 2’s proof under which the concavity of the prosumer’s objective function is guaranteed. However, it is worth noting that $\mathcal{A} \subset \mathcal{C}$ where $\mathcal{C}$ is the set defined by all possible $\theta_i$'s in which the objective function \eqref{eq16a} is concave in $\theta_i$.

Beside the region of concavity of \eqref{eq16a} in $q_i$, defined above in \eqref{eq36}, we want to show that $\tilde{S_i}(q_i), i \in \mathcal{N}$ (hence \eqref{eq16a}) is strictly concave in $q_i$. To show this, let $q_i,q_i' \in \mathbb{R}$ where $-s_{max} \leq q_i < q_i'$. Then, $\tilde{S_i}(q_i)$ is strictly concave in $q_i$ when $\frac{\partial \tilde{S_i}(q_i')}{\partial q_i'} < \frac{\partial \tilde{S_i}(q_i)}{\partial q_i}$. The first derivative of $\tilde{S_i}(q_i)$ is given by:
\begin{equation}\label{eq37}
\frac{\partial \tilde{S_i}(q_i)}{\partial q_i} = (1 + \frac{q_i}{(N-1)d_{min}}) \frac{\partial S_i(q_i)}{\partial q_i}
\end{equation}
From \eqref{eq37} and by Assumption 2, it is easy to see that $\frac{\partial \tilde{S_i}(q_i')}{\partial q_i'} < \frac{\partial \tilde{S_i}(q_i)}{\partial q_i}$. Hence, \eqref{eq16a} is strictly concave. Under the above conditions and the fact that $\tilde{S_i}(q_i)$ is continuous in $q_i$, there exists a unique market allocation profile $\boldsymbol{\tilde{q}}$.

\textbf{Necessary and Sufficient Conditions for the Market Allocation:}
\vspace{0.1cm}

Given that the Slater’s constraint qualification holds for \eqref{eq16}, then the KTT optimality conditions are necessary and sufficient. Associate the Lagrange multipliers $\eta$ with the equality constraint and $\lambda_i, i \in \mathcal{N}$ with the inequality constraints. The Lagrangian function becomes:
\vspace{-0.1cm}
\small \begin{equation} \label{eq38}
L(\boldsymbol{q},\eta, \boldsymbol{\mu}) = \sum_{i=1}^N \tilde{S_i}(q_i) - \eta \Big( \sum_{i=1}^N q_i\Big) + \sum_{i=1}^N \lambda_i (q_i + s_{max})
\vspace{-0.2cm}
\end{equation}\\
\normalsize
Let $\tilde{\boldsymbol{q}}$ be the unique optimal solution to \eqref{eq16}. Analyzing the KKT conditions shows that $\tilde{\boldsymbol{q}}$ and $\tilde{\eta}$ must satisfy:
\vspace{-7pt}
\begin{subequations}\label{eq39}
\begin{equation}\label{eq39a}
 \frac{\partial S_i(\tilde{q_i})}{\partial \tilde{q_i}} (1 + \frac{\tilde{q_i}}{(N-1)d_{min}})  = \tilde{\eta}, \ \text{if} \ \tilde{q_i}>-s_{max}
\end{equation}
\begin{equation}\label{eq39b}
 \frac{\partial S_i(\tilde{q_i})}{\partial \tilde{q_i}} (1 + \frac{\tilde{q_i}}{(N-1)d_{min}})  < \tilde{\eta}, \ \text{if} \ \tilde{q_i}=-s_{max}
\end{equation}
\end{subequations}

\textbf{Uniqueness of the Nash Equilibrium:}
\vspace{-0.00cm}

Analogous to the analysis in the last paragraph of the proof of Theorem 1, it is straightforward to conclude the one-to-one correspondence of \eqref{eq34} and \eqref{eq39}. Therefore, uniqueness of the Nash equilibrium follows from the corresponding unique market allocation. That is, uniqueness of the Nash equilibrium follows from the one-to-one correspondence of $(\boldsymbol{\tilde{q}}, \tilde{\eta})$ to $\boldsymbol{\tilde{\theta}}$ with $\tilde{\eta}=p(\boldsymbol{\tilde{\theta}})$.
\end{appendices}


\end{document}